\documentclass[journal]{IEEEtran}
\ifCLASSINFOpdf
\else
\fi
\usepackage{booktabs} 
\usepackage{amssymb}
\usepackage{graphicx}
\usepackage{caption}
\usepackage{subfigure}
\usepackage{algorithm}  
\usepackage{algorithmic}  
\usepackage{multirow}
\usepackage{booktabs}
\usepackage{amsmath}
\usepackage{url}
\usepackage {float}
\usepackage{diagbox}
\renewcommand{\baselinestretch}{1.059}
\begin{document}

%
\title{Catfish Effect Between Internal and External Attackers:Being Semi-honest is Helpful}
%
%
%

\author{Hanqing~Liu,~\IEEEmembership{Student Member,~IEEE,}
        Na~Ruan,~\IEEEmembership{Member,~IEEE,}
        and~Joseph~K.~Liu,~\IEEEmembership{Member,~IEEE}

\thanks{Hanqing Liu is with Shanghai Jiaotong University, Shanghai, P.R. China}
\thanks{Na Ruan is with MoE Key Lab of Artificial Intelligence, Shanghai Jiaotong University, Shanghai, P.R. China. Na Ruan is the corresponding author (e-mail: naruan@cs.sjtu.edu.cn)}
\thanks{Joseph K.Liu is with Monash University, Melbourne, Australia}
}

\maketitle

\begin{abstract}
The consensus protocol named proof of work (PoW) is widely applied by cryptocurrencies like Bitcoin. Although security of a PoW cryptocurrency is always the top priority, it is threatened by mining attacks like selfish mining. Researchers have proposed many mining attack models with one attacker, and optimized the attacker's strategy. During these mining attacks, an attacker pursues a higher relative revenue (RR) by wasting a large amount of computational power of the honest miners at the cost of a small amount of computational power of himself. 

In this paper, we propose a mining attack model with two phases: the original system and the multi-attacker system. It is the first model to provide both theoretical and quantitative analysis of mining attacks with two attackers. We explain how the original system turns into the multi-attacker system by introducing two attackers: the internal attacker and the external attacker. If both attackers take the attacking strategy selfish mining, the RR of the internal attacker in multi-attacker system will drop by up to 31.9\% compared with his RR in original system. The external attacker will overestimate his RR by up to 44.6\% in multi-attacker system. Unexpected competitions, auctions between attackers and overestimation of attackers' influence factor are three main causes of both attackers' dropping RR. We propose a mining strategy named Partial Initiative Release (PIR) which is a semi-honest mining strategy in multi-attacker system. In some specific situations, PIR allows the attacker to get more block reward by launching an attack in multi-attacker system.
\end{abstract}

%
\IEEEpeerreviewmaketitle

\section{Introduction}
%
%
%
%
\IEEEPARstart{N}{owadays}, most traditional payments on the Internet are based on trusted third parties. The trust based payment model has some shortcomings which make completely non-reversible transactions impossible \cite{2}. In the past decade, public's interest has focused on decentralized cryptocurrencies based on cryptographic proof. These cryptocurrencies, presented by Bitcoin which is the first fully decentralized cryptocurrency \cite{21}, rely on blockchain technology to guarantee their security. Generally speaking, a blockchain is an open ledger used to store and maintain a list of record. Its consensus protocol guarantees that previous data written in a blockchain is irreversible and consistent to all users. A series of consensus protocol such as proof of work (PoW), proof of stake (PoS) and practical Byzantine fault tolerance (PBFT) are applied to cryptocurrencies. Proof of work, applied by Bitcoin, takes the largest market share. In 2016, proof-of-work blockchains take about 90\% of the market \cite{9}. In a proof-of-work blockchain such as Bitcoin, the participants are required to generate PoWs by solving cryptographic puzzles when they are finding blocks. The one who tries to generate new blocks are called miners, and the process to find a new block is called mining. When a miner finds a new block, he propagates his solution to other miners. All other miners confirm his solution and beginning solving a new crytographic puzzle  \cite{2}. To lower the variance of a solo miner's revenue, a group of miners are organized to form a mining pool.

A miner or a mining pool with a large amount of computational power may threaten the blockchain. The most basic requirement for a proof-of-work blockchain is that the records in the blockchain cannot be modified. But a well-known attacking strategy named 51\% attack can revert the transaction records in a proof-of-work blockchain \cite{2,12}. 51\% attack requires the attacker to control more than 50\% of the network's computational power. In 2018, a proof-of-work cryptocurrency called Bitcoin Gold suffered 51\% attack. 51\% attack requires a large amount of computational power and breaks the reliability of a blockchain. Generally speaking, only those attackers whose aim is to revert transaction records will launch a 51\% attack. 

 For these attackers, block reward can be one of their aims. In a proof-of-work blockchain, the ideal case is that a miner or a mining pool with $\alpha$ fraction of computational power in the blockchain should gain $\alpha$ fraction of block reward in a long period of time. But many studies have indicated that an attacker in a proof-of-work blockchain can get more share of block reward than he deserves through some strategies. The most well-known one is selfish mining presented by Eyal and Sirer in 2014 \cite{1}. In May 2018, a Japanese proof-of-work cryptocurrency was hit by selfish mining attack and the attack has caused roughly \$90,000 in damages. 

Like Bitcoin Gold and Monacoin, Bitcoin is a proof-of-work based powered blockchain as well. A statistic from \textit{blockchain.info} indicates that, in 2017, the mining difficulty which is approximately proportional to the whole computational power in Bitcoin has increased by four times. The huge computational power in Bitcoin makes it unrealistic to launch a 51\% attack to Bitcoin. But the threshold of the computational power to launch a selfish mining attack is far lower. Another statistic from \textit{blockchain.info} demonstrate that, in 2018, the price of Bitcoin has dropped by more than 70\%. The dropping price may result in the fleeing of computational power in Bitcoin which can lower the threshold of computational power to launch a selfish mining attack. Under this situation, Bitcoin has to face the fact that it might suffer from selfish mining attack in the future. Many former studies \cite{4,8,21} have presented attacking strategis towards a proof-of-work blockchain which have better performance than selfish mining. They put their emphasis on the optimization of one single attacker's strategy.


Former studies do not consider the case with two or more attackers. We assumes that the number of attackers may increase to two. In this paper, we establish a mining attack model of a proof-of-work blockchain with two attackers. To the best of our knowledge, this is the first effort on systematically modeling a mining attack model with two attackers. We explain why the second attacker may occur by dividing attackers into two types: the internal attacker and the external attacker. We define the proof-of-work cryptocurrency system which consists of the honest miner and internal attacker as original system, and the system which consists of the honest miner, the internal attacker and the honest miner as multi-attacker system. Our work reveals that the internal attacker weakens the original system, so the original system is more likely to be attacked by the external attacker. And in multi-attacker system, both external attacker and internal attacker's revenue do not meet their expectation. This paper makes the following contributions:

\textbf{Contribution 1: Establishing a proof-of-work blockchain model with internal attacker and external attacker.} We propose a new proof-of-work blockchain model, which consists of two attackers. The decision making process, the mining behavior, the state transition and the attacking strategies of the attackers are different from the former blockchain models with one single attacker.

\textbf{Contribution 2: Theoretical and quantitative analysis on the conditions that external attacker may occur.} Our analysis reveals the relationship between the computational power of the internal attacker and the degree that the original system's computational power is weaken. We prove that the external attacker may attack the original system if his computational power is in a certain scope.

\textbf{Contribution 3: A new attacking strategy named Partial Initiative Release (PIR).} We have proved that both the internal attacker and the external attacker may face the fact that they do not gain as much revenue as their expectation in the multi-attacker system. We discuss the Catfish Effect in the multi-attacker system. We propose a new attacking strategy called Partial Initiative Release (PIR) which is the countermeasure for the internal attacker after he notices the existence of the external attacker. Our simulations demonstrate that internal attacker with strategy PIR can gain more revenue than the external attacker with strategy selfish mining.



The previous version of this paper has appeared in \cite{18}. This paper has extended and improved the previous version. The most important extensions includes distinguishing the type of attackers and the phase of our mining attack model (Section \uppercase\expandafter{\romannumeral3}. A), theoretical analysis and quantitative analysis on how internal attacker weakens the original system (Section \uppercase\expandafter{\romannumeral4}), theoretical analysis and quantitative analysis on the condition that the external attacker may attack the original system (Section \uppercase\expandafter{\romannumeral5}), discussion on the Catfish Effect in multi-attacker system (Section \uppercase\expandafter{\romannumeral6}. C). Compared with the previous version \cite{18}, distinguishing the type of attackers and the phase of our mining attacker model better explains why and how two attackers occur, newly added theoretical and quantitative analysis provides more solid proofs, and the discussion on Catfish Effect explains more clearly why the attacks need to switch their attacking strategy in multi-attacker system.

The rest of our paper is organized as follows: We begin by introducing the basic concepts of proof-of-work blockchain in Section \uppercase\expandafter{\romannumeral2}. In Section \uppercase\expandafter{\romannumeral3} we introduce our proof-of-work blockchain model including attackers' state space, action space and assumptions in our analysis. In Section \uppercase\expandafter{\romannumeral4} we present the theoretical and quantitative analysis about how the internal attacker weakens the original system. In Section \uppercase\expandafter{\romannumeral5}, we provide with the condition that the external attacker may attack the original system. We also illustrate the fact that the external attacker will fail to gain as much revenue as he expects. In Section \uppercase\expandafter{\romannumeral6}, we describe the Catfish Effect in multi-attacker system and propose the PIR strategy for the internal attacker. In Section \uppercase\expandafter{\romannumeral7}, we conclude our paper.

\section{Preliminaries}

In this section, we describe some basic concepts of Bitcoin and some attacking strategies in Bitcoin including selfish mining since Bitcoin is the most well-known and most typical instance of proof-of-work blockchain. The basic concepts and attacking strategies are also suitable for other proof-of-work blockchain instances.

\subsection{Basis of Bitcoin }

\subsubsection{Miner and Mining Process}
In Bitcoin, the miners are the participants who are working on finding new blocks. The process for the miners to find new blocks are called mining process. In the blockchain of Bitcoin, the block header identifies each block \cite{2}. A block header consists of the hash of previous block header, the Merkle root of the transactions stored in the block and a nonce. The miners' work is to select the transactions which has not been stored in previous blocks can generate nonces. If the hash value of all data in the block header is lower than a specific threshold t, then the miner can propagate the new block found by him to the Bitcoin network. Other miners will accept the new block after verification. The mining process of the miner seriously rely on the miner's computational power. In Bitcoin, the threshold t is adjusted about every two weeks \cite{2}. The more computational power is in the Bitcoin system, the lower the threshold is and the more difficult it is to find new blocks. 

For a solo miner, he has to wait for a long time which is not intolerable before he finds a new block \cite{21}. To gain block revenue in a more stable way, a group of solo miners organize a mining pool. Mining pools benefit their members, but increases uncertainty to the whole system. Once the pool manager of a mining pool wants to gain more revenue than the share he deserves, with a large amount of computational power, he can easily launch mining attacks. The largest pool in Bitcoin history has the computational power which exceeds 40\% of the computational power in Bitcoin system \cite{8}.

\subsubsection{Honest Miners and Attackers}
Honest miners in Bitcoin follow Bitcoin protocol. They immediately propagate their newly found blocks to the Bitcoin network and accept the longest blockchain as their main chain. If there exists a fork in blockchain, the honest miners accept the block they receive first. 

Solo miners and mining pools can be the attackers in Bitcoin and act in a different way. Some typical attacking tricks are \cite{1,3,5,8,14}:
\begin{itemize}
    \item[1)] Denial of propagating the blocks which is found by others to other miners.
    \item[2)] Denial of immediately propagating the blocks found by themselves to the Bitcoin network.
    \item[3)] Denial of accepting the longest chain instead of their own chain as the main chain.
\end{itemize}

The behavior of the attackers includes but is not limit to the tricks above. An attacker can adjust their behavior and choose which tricks to be used according to his attacking strategy.

\subsubsection{Forks in Blockchain}
A fork in Bitcoin occurs when two miners find and propagate their newly found block at roughly the same time. Due to the information propagation delay in Bitcoin network \cite{3,15}, part of honest miners receive and verify one branch of the fork first while other honest miners receive and verify the other branch of the fork. This kind of forks is randomly generated and will be eliminated when the next block is found so that one branch is extended and accepted by all the honest miner while the other branch is staled. Gervais \cite{7} and Decker \cite{3} estimate the probability that a randomly generated fork occurs in Bitcoin. Both two works suggest that the randomly generated fork rate of Bitcoin ranges from 0.41\%  \cite{7} to 1.7\% \cite{3} according the information propagation in Bitcoin.

An attacker can generate a fork intentionally. In mining attacks including selfish mining \cite{1,4,8,21}, attacker  will intentionally generate a fork and provoke a competition in the blockchain. Intentionally generated forks always means the waste of computational power.

\subsection{Mining Attacks}
\subsubsection{Selfish Mining}
Bitcoin, as the proof-of-work cryptocurrency, is designed under the assumption that as long as the majority of the hashpower is honest, Bitcoin's safety is guaranteed \cite{2}. But this assumption has been overthrown by selfish mining proposed by Eyal and Sirer \cite{1}. Selfish mining allows the adversarial miners or mining pools to get more revenue than they deserve. The attackers do not immediately release the blocks found by themselves. They do not accept  the longest chain as their main chain as long as they are holding some unreleased blocks. An attacker with 33\% of the computational power of Bitcoin system can ensure that he can earn extra revenue (more than 33\% of the revenue of the entire system). The threshold can even lower to 0 with the increase of information propagation delay among the honest miners.

\subsubsection{Optimization of Selfish Mining}
After Eyal and Sirer's work, many researchers are focusing on optimization of selfish mining. Sapirshtein's work \cite{4} and Nayak's work \cite{8} extend the attackers' strategy space. Nayak's work \cite{8} also combines selfish mining and eclipse attack \cite{6} which is a network-level attack. Kwon's work \cite{21} combines selfish mining with block withholding attack \cite{16}. The former studies present two possible approaches to optimize selfish mining: Extending the attacker's strategy space or combining selfish mining with other mining attacks. The former works only consider the attacking scenes with a single attacker. 

\subsubsection{Methods to Evaluate a Strategy}
Gervais's work points out that selfish mining is an irrational strategy in a short term since it wastes both the attacker's and honest miner's computational power \cite{7}. But in a long term, Bitcoin will lower the mining difficulty \cite{2}. So for selfish mining and mining attacks which optimize selfish mining, block reward can not directly measure the performance of the attacking strategies. In a short term the attackers' aim is increasing their fraction of block reward of the entire system instead of increasing block reward directly. According to attacker's aim, relative revenue (RR) and stale block rate (SBR) are used to evaluate the performance of a attacking strategy in many former works \cite{1,4,5,8}.

\section{Attack model}

In this section, we introduce our attack model from the following aspects: two phases of our model, attackers' state and action , attackers' decision making process and the evaluation of attackers' revenue.

\subsection{Two Phases of Our Model}
There are two attackers in our model: the internal attacker and the external attacker. Either a solo miner or a mining pool can act as an attacker. The honest miners, no matter whether they are solo miners or mining pools, accept the same main chain when there are no forks in the blockchain. When no forks exist, the honest miners can be seen as an whole honest entity. Otherwise, the computational power of the honest miners splits due to information propagation delay.

We define the first phase of our model as the original system. The original system consists of the internal attacker and the honest miner. In the original system, internal attacker can launch a selfish mining attack. After the attack, the original system can be considered as a selfish mining model with one single attacker. 

The second phase of our model is defined as the multi-attacker system. The multi-attacker system consists of the internal attacker, the external attacker and the honest miner. The multi-attacker system results from the external attacker's selfish mining attack against original system after internal attacker's attack.

\subsection{State and Action}
\subsubsection{Attackers' state}
Each attacker's state contains the information of the attacker in the blockchain. The attackers make decisions based on their state. Our attacking model, with two attackers, considers some special states which do not exist in those models with a single attacker. The following information should be included in the state:
\begin{itemize}
    \item[1)] The attacker's lead: The private chain of an attacker consists of the main chain accepted by the attacker and the unreleased blocks. We define an attacker's lead as the height of the attacker's private chain minus the main chain accepted by the honest miner.
    \item[2)] Whether the attacker is in a competition or not: When an attacker intentionally generates forks in Bitcoin, it is possible for him to be involved in a competition. Whether the attacker is in a competition or not determines his next action.
    \item[3)] If there are another fork in the blockchain: Another fork means the fork which is not generated by the attacker randomly or intentionally. If other miners release a new block at roughly the same time, there will be an competition in the blockchain which the attacker is not engaged in.
\end{itemize}

We use the notation in selfish mining attack model to represent the attacker's state. $S = i~(i = 0,1,2\ldots)$ means that the attacker's lead is $i$ and there is no forks in the blockchain. $S = 0'$ represents that the attacker is in a competition with other miners.

The notation of the attacker's state above is designed for the attacking model with a single attacker. It cannot cover all situations in our model. We define $S =i''~(i = 0,1,2\ldots)$ to cover the situations that there is a fork in the blockchain, but the attacker is not involved in the competition.

\subsubsection{Attacker's action}
Attacker's action determines whether the attacker should release his blocks or not and how the attacker release his blocks. Similar to attacker's state, we use the notation of attacker's action in selfish mining attack model. But the meaning of notations is adjusted so as to be suitable for our model with two attackers. The attacker have five basic actions:

Hold: The attacker do not release any blocks or select a new main chain.

Match: The attacker releases one or $n$ of his unreleased blocks so that the attacker's released chain can catch up the other miner's chain.
Override: The attacker releases two or $n$ of his unreleased blocks so that the attacker's released chain can exceed the other miner's chain just right.

Adopt: The attacker gives up on his private chain and select the longest chain as his main chain. 

Release: The attacker extend his released chain by one block.
\subsection{Decision Process}

An attacker needs to decide which basic action he should take and when to take the basic action based on his state. The whole process is decision process of the attacker.

Any attacker faces a Markov decision problem: $M = (S,A,P,R)$ where $S$ is attacker's state, $A$ is attacker's action space which consists of five basic actions, $P$ is the probability of attacker's state transition and $R$ is the revenue of attacker's action. If we denote the attacker's previous state as $S_{prev}$ and attacker's current state as $S_{cur}$, then we have the state transition equation for the attacker at any state:
\begin{equation}
    P_a(i,j) = P(S_{cur} = j~|~ S_{prev} = i~, ~Action_{prev} ~ = a)
\end{equation}

The processing of finding the best response in the next $n$ steps is usually too complicated for the attacker. Thus an attacker needs to apply a specific mining attack strategy (Expressed in the form of a state machine) to find a sub-optimal response. A mining attack strategy can be considered as a method to reduce the complexity of finding a solution at the cost of part of the revenue.

\subsection{Attacker's Revenue}
Attacker's revenue is used to quantify whether the attacker can gain extra revenue from his attack. Our model consider all mining attacks as irrational so that block reward is not suitable for our model. When quantifying the attacker's or the honest miner's revenue, we use stale block rate (SBR) and relative revenue (RR). 

A miner's SBR shows how much computational power of the miner is wasted. Denote the number of blocks which are found by the miner and accepted by all honest miners as $Num_{ac}$. And denote the number of blocks which are found by the miner and not accepted by all honest miners as $Num_{nac}$. The miner's SBR can be calculated as:
\begin{equation}
    SBR = \frac{Num_{ac}}{Num_{ac}+Num_{nac}}
\end{equation}

A miner's RR shows whether a miner receives as much block reward as he deserves. Denote the number of blocks which are found by other miners and accepted by all honest miners as $Num_{oac}$. The miner's RR can be calculated as:
\begin{equation}
    RR = \frac{Num_{ac}}{Num_{ac}+Num_{oac}}
\end{equation}

\subsection {Our assumptions}
We have made the following five assumptions in our analysis:
\begin{itemize}
\item[1)] There are three participants in our model: The honest miner, the internal attacker and the external attacker. The internal attacker mines in the original system from the beginning, and he decides to starts selfish mining attack. The external attacker joins into the system after the internal attacker's attack. 

\item[2)] The honest miner, the internal attacker and the external attacker can either be a solo miner or a mining pool. We make the assumption that honest miner's computational power is always greater than the internal attacker's and the external attacker's.
\item[3)] We do not consider the mining attack strategy which do not intentionally create forks, such as Eclipse attack and block withholding attack. 

\item[4)] The total computational power of the honest miner, the internal attacker and the external attacker is normalized which means that the total computational power of the multi-attacker system is 1. Meanwhile, we assume that the total computational power will not change any more after the external attacker's participation.

\item[5)] Both the internal attacker and the external attacker are selfish mining attackers at the beginning. After noticing the existing of the external attacker, the internal attacker will take some countermeasures. During their attack, except the forks intentionally created by them, we do not consider the randomly generated forks since the fork rate is negligible.
\end{itemize}

\section{Original System}
In this section, we will demonstrate how the internal attacker's attack weakens the computation power of the original system which consists of the honest miner and the internal attacker from two aspect: Theoretical analysis and simulation. In this section, the internal attacker launches a selfish mining attack.
\subsection{Theoretical Analysis}
The relevant parameters are as follows:
\begin{itemize}
\item $\alpha$: Computational power of the honest miner the honest miner.
\item $\beta_1$: Computational power of the internal attacker.
\item $\beta_2$: Computational power of the external attacker.
\item $\tau_1$: Probability that the internal attacker's chain win the competition when the internal attacker is competing with the honest miner or the external attacker.
\item $\tau_2$: Probability that the external attacker's chain win the competition when the external attacker is competing with the honest miner or the internal attacker.
\item $\gamma_1$: The fraction of honest miner that helps the internal attacker when the internal attacker's chain is competing with others.
\item $\gamma_2$: The fraction of honest miner that helps the external attacker when the external attacker's chain is competing with others.
\item $\gamma_h$: The fraction of honest miner that helps the attackers when there is a competition.

\end{itemize}

First, we normalize the computational of the original system first. In the original system, the fraction of the computational power of internal attacker is: $\beta' = \frac{\beta_1}{\beta_1+\alpha}$, and the fraction of computation power of the honest miner is: $\alpha' = \frac{\alpha}{\beta_1+\alpha}$. 

\begin{figure}[htbp]
    \centering
    \includegraphics[width = 3in]{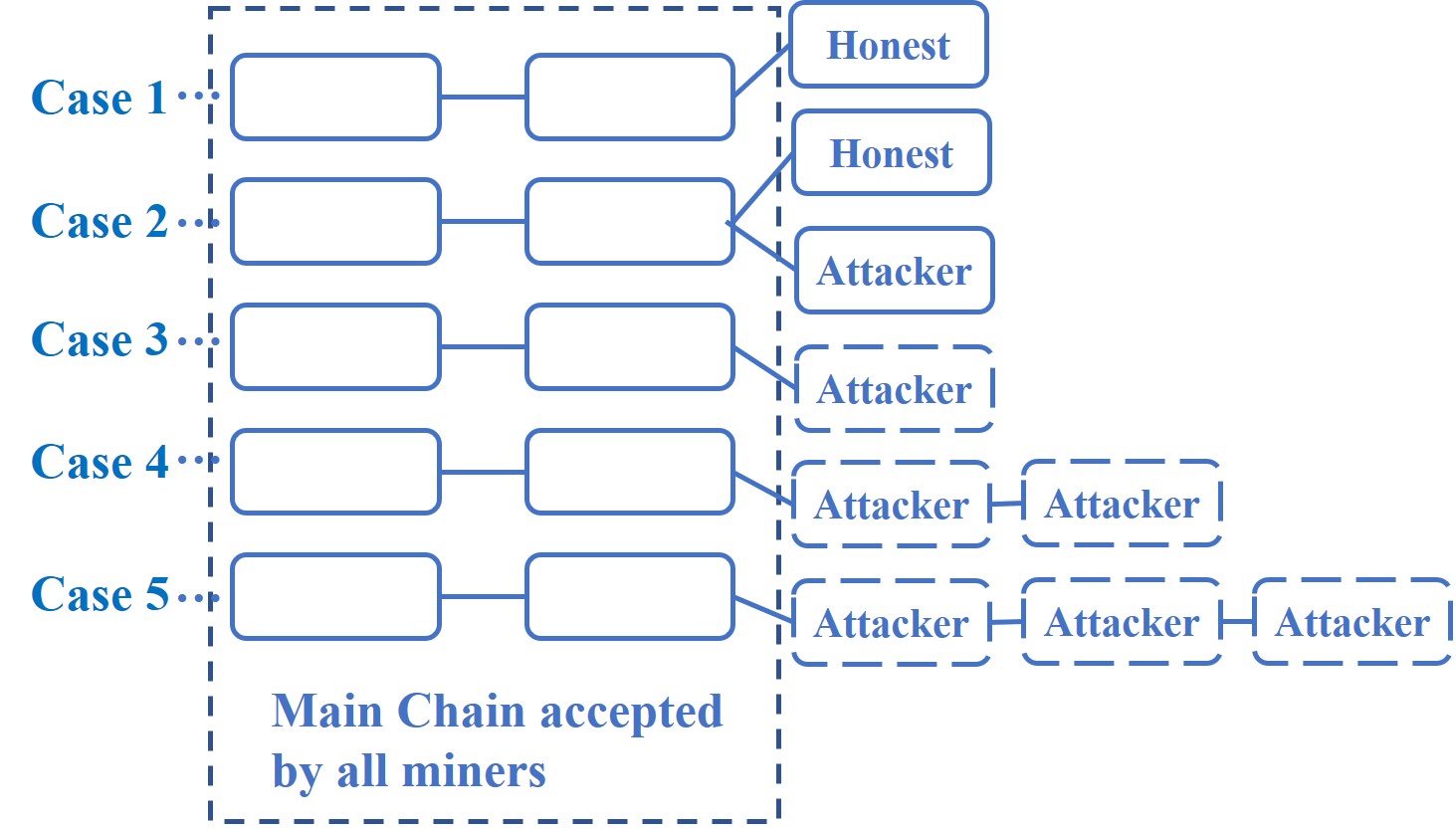}
    \caption{5 cases}
    \label{5cases}
\end{figure}

To show how the internal attacker weakens the original system, we classify the case based on the internal attacker's state as shown in Fig.\ref{5cases}. In the first case, the internal attacker is at state 0 with the probability $Pr(S_{\beta_1} = 0)$. In the second case, the internal attacker is at state 0' with the probability $Pr(S_{\beta_1} = 0')$. In this case, the internal attacker has the probability $\tau_1$ to win the competition, and the probability that the internal attacker finds the next block in the competition is $\beta'$. Note that, $\beta'$ is not necessarily equal to $\tau_1$. Typically, due to the information propagation delay and some other factors, part of the honest miner's computational power will help the internal attacker in the competition, which means that $\tau_1$ is usually greater than $\beta'$. In the third case, the internal attacker is at stage 1 with the probability $Pr(S_{\beta_1} = 1)$.In the forth case, the internal attacker is at state 2 with the probability $Pr(S_{\beta_1} = 2)$. In the final case, the internal attacker has the state that greater than two with the probability $Pr(S_{\beta_1} \ge 3)$.

\newtheorem{lemma}{Lemma}[section]
\begin{lemma}\label{lm1}
The probability that the internal attacker is at any state is:
\end{lemma}
\begin{equation}
\left\{
\begin{array}{lr}
Pr(S_{\beta_1} = 0) = \frac{\beta'-2\beta'^2}{\beta'(2\beta'^3 - 4\beta'^2 + 1)},&\\
Pr(S_{\beta_1} = 0') = \frac{\alpha'(\beta'-2\beta'^2)}{2\beta'^3 - 4\beta'^2 + 1},&\\
Pr(S_{\beta_1} = k) = (\frac{\beta'}{\alpha'})^{k-1}\frac{\beta'-2\beta'^2}{2\beta'^3 - 4\beta'^2 + 1},&For~ k \ge 1
\end{array}
\right.
\end{equation}

\begin{IEEEproof}
From the selfish mining state machine, we can focus on one unique point: $S_{\beta_1} = 2$. When  $(S_{\beta_1} \ge 2)$, the internal attacker's state transition probability from state $k$ to state $k+1$ is $\beta'$ which is always less than 0.5. Thus, for $k \ge 2$, we have $\beta'Pr(S_{\beta_1} = k) = \alpha'Pr(S_{\beta_1} = k+1)$. In addition, based on a selfish miner's behavior when his state is less than 2, we can derive:

\begin{equation}
\left\{
\begin{array}{lr}
Pr(S_{\beta_1} = 0) = Pr(S_{\beta_1} = 0')+ (\alpha')Pr(S_{\beta'} = 2),&\\
Pr(S_{\beta_1} = 0') = \alpha'Pr(S_{\beta_1} = 1),&\\
Pr(S_{\beta_1} = 1) = \alpha'Pr(S_{\beta_1} = 2),&\\
\beta'Pr(S_{\beta_1} = k) = \alpha'Pr(S_{\beta_1} = k+1),& k \ge 2\\
\sum_{k=0}Pr(S_{\beta_1} = k) + Pr(S_{\beta_1} = 0') = 1
\end{array}
\right.
\end{equation}
With equation(2), we can finally derive equation(1) in Lemma 1. Further, the probability of the five cases we discuued above is:
\begin{equation}
\left\{
\begin{array}{lr}
Pr(Case 1) = \frac{\beta'-2\beta'^2}{\beta'(2\beta'^3 - 4\beta'^2 + 1)},&\\
Pr(Case 2) = \frac{\alpha'(\beta'-2\beta'^2)}{2\beta'^3 - 4\beta'^2 + 1},&\\
Pr(Case 3) = \frac{\beta'-2\beta'^2}{2\beta'^3 - 4\beta'^2 + 1},&\\
Pr(Case 4) = \frac{\beta'(\beta'-2\beta'^2)}{\alpha'(2\beta'^3 - 4\beta'^2 + 1)},&\\
Pr(Case 5) = \frac{\beta'^2}{\alpha'(\alpha'-\beta')}\frac{\beta'-2\beta'^2}{2\beta'^3 - 4\beta'^2 + 1},&\\
\end{array}
\right.
\end{equation}
\end{IEEEproof}

\begin{lemma}\label{lm2}
The probability that the internal attacker can extend the length of chain when a new block is found is: $\frac{\beta'(\beta'-2\beta'^2)}{2\beta'^3 - 4\beta'^2 + 1} + \frac{\beta'\alpha'(\beta'-2\beta'^2)}{2\beta'^3 - 4\beta'^2 + 1}$. It is irrelevant to the value of $\gamma_1$.
\end{lemma}

\begin{IEEEproof}
In Case 1, the internal attacker has the probability $1-\beta'$ to remain his state 0 and has the probability $\beta'$ to move on to state 1. In Case 3, the internal attacker has the probability $1-\beta'$ to state 0' and has the probability $\beta'$ to state 2. In Case 5, the internal attacker has the probability $1-\beta'$ to state $k-1$ and has the probability $\beta'$ to state $k+1$. None of the six results shown above can result in the increase of the main chain in Bitcoin. 

In Case 2, the internal attacker has the probability $\tau_1 = \beta' + \gamma_1 \alpha' $ to win the competition and increase the length of main chain by 1. The probability that the competition is won by the honest miners who support the internal attacker is $\frac{\tau_1 - \beta'}{\tau_1}$. So the probability that the internal attacker wins the competition can increase the length of main chain by himself is actually $\frac{\beta'\alpha'(\beta'-2\beta'^2)}{2\beta'^3 - 4\beta'^2 + 1}$. In Case 4, the internal attacker has the probability $\beta'$ to increase his state to 3 which will not result in the increase of the main chain. So, the probability that the internal attacker increase the length of main chain in this case is $\frac{\beta'(\beta'-2\beta'^2)}{2\beta'^3 - 4\beta'^2 + 1}$.

From the analysis above, The probability that the internal attacker can lengthen the length of chain when a new block is found by original system is: $\frac{\beta'(\beta'-2\beta'^2)}{2\beta'^3 - 4\beta'^2 + 1} + \frac{\beta'\alpha'(\beta'-2\beta'^2)}{2\beta'^3 - 4\beta'^2 + 1}$
\end{IEEEproof}
\begin{lemma}\label{lm3}
The original system's probability to extend the main chain when a new block is found is always less than 1.
\end{lemma}
\begin{IEEEproof}
For the honest miner, the probability to extend the main chain is always $\alpha'$, so the probability that the original system can extend the main chain when a new block is found is: $\alpha' + \frac{\beta'(\beta'-2\beta'^2)}{2\beta'^3 - 4\beta'^2 + 1} + \frac{\beta'\alpha'(\beta'-2\beta'^2)}{2\beta'^3 - 4\beta'^2 + 1}$. Let $f(\beta') = \alpha' + \frac{\beta'(\beta'-2\beta'^2)}{2\beta'^3 - 4\beta'^2 + 1} + \frac{\beta'\alpha'(\beta'-2\beta'^2)}{2\beta'^3 - 4\beta'^2 + 1}-1$. \\
$\frac{\mathrm{d}f(\beta')}{\mathrm{d}\beta'} = \frac{-8\beta'^4+ 9\beta'^2-4\beta'}{4\beta'^6 - 16\beta'^5-16\beta'^4+4\beta'^3-8\beta'^2+1}\le 0$ when $\beta'$ ranges from 0 to $\frac{1}{2}$. $f(\beta') \le f(0) = 0$. Then, we can derive the inequality:$
 \alpha' + \frac{\beta'(\beta'-2\beta'^2)}{2\beta'^3 - 4\beta'^2 + 1} + \frac{\beta'\alpha'(\beta'-2\beta'^2)}{2\beta'^3 - 4\beta'^2 + 1} \le 1$

\end{IEEEproof}

Generally speaking, after the selfish mining attacker launched by the internal attacker, in a long period of time(in Bitcoin, about 2 weeks), the computational power of the original system is equivalent to a single honest miner with the computational power $(\alpha + \beta_1)(\alpha'+ \frac{\beta'(\beta'-2\beta'^2)}{2\beta'^3 - 4\beta'^2 + 1} + \frac{\beta'\alpha'(\beta'-2\beta'^2)}{2\beta'^3 - 4\beta'^2 + 1})$. The factor $\rho = (\alpha'+ \frac{\beta'(\beta'-2\beta'^2)}{2\beta'^3 - 4\beta'^2 + 1} + \frac{\beta'\alpha'(\beta'-2\beta'^2)}{2\beta'^3 - 4\beta'^2 + 1})$ shows the degree that internal attacker the internal attacker weakens the original system. We name the factor $\rho$ as the shrinkage factor.

\subsection{Quantitative Analysis and Simulation}
Theoretically, from the external attacker's perspective, the original system's computational power will shrink by  $\rho$. We use the definition stale block rate(SBR) to show how much computational power the internal attacker and honest miner have lost after the internal attacker launches selfish mining attack in this simulation. We use a Monte Carlo method to generate a blockchain with the height $10^6$ blocks by 100 times.

In this simulation, we consider the a simple case: The external attacker has not joint the whole system yet and the internal attacker launches a selfish mining attack to the honest miner. 

\begin{figure}[htbp]
\centering
\subfigure[Stale block rate]{
\begin{minipage}[t]{0.5\linewidth}
\centering
\includegraphics[width=1.6in]{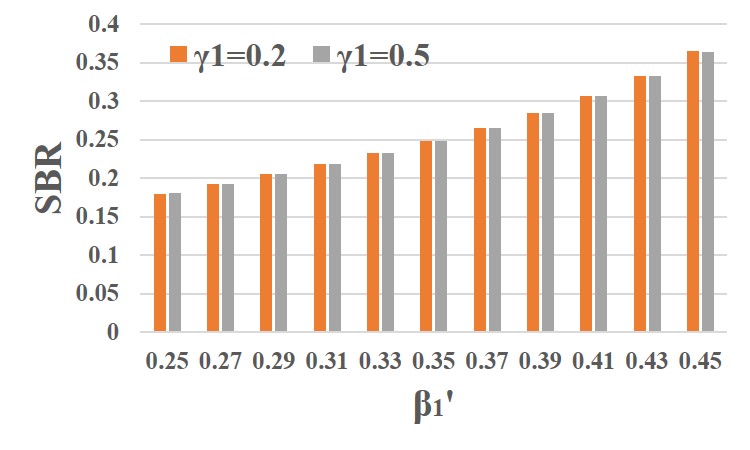}
\end{minipage}%
}%
\subfigure[The shrinkage factor]{
\begin{minipage}[t]{0.5\linewidth}
\centering
\includegraphics[width=1.6in]{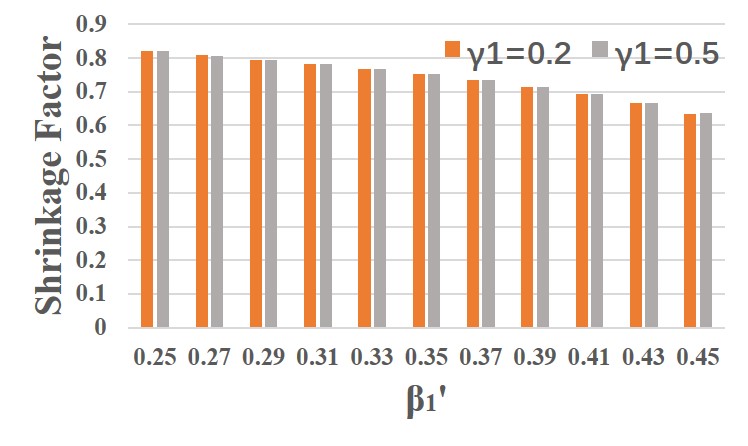}
\end{minipage}%
}%

\centering
\caption{The simulation result of the original system}
\label{shrink}
\end{figure}

Fig. \ref{shrink}(a) shows the SBR of the original system, given the internal attacker's computational power, when the parameter $\gamma_1$ is 0.2 and 0.5 respectively. Fig. \ref{shrink}(a) demonstrates that when the computational power of the internal attacker(normalized) increases, the stale block rate of the original system also increases. The simulation result in Fig. \ref{shrink}(a) also indicates that the SBR of the original system is irrelevant to the parameter $\gamma_1$. This result confirms \textit{Lemma 2}

Fig. \ref{shrink}(b) shows the shrinkage factor in the simulation. In this simulation, shrinkage factor is equal to $1-SBR$. The more computational power is wasted due to the internal attacker's attack, the less shrinkage factor is. 

			
			
		

Besides, in theoretic the shrinkage factor can be calculated as $\rho = (\alpha'+ \frac{\beta'(\beta'-2\beta'^2)}{2\beta'^3 - 4\beta'^2 + 1} + \frac{\beta'\alpha'(\beta'-2\beta'^2)}{2\beta'^3 - 4\beta'^2 + 1})$ where $\beta' = \frac{\beta_1}{\beta_1+\alpha}$, and $\alpha' = \frac{\alpha}{\beta_1+\alpha}$. Table \ref{tab1} compares the shrinkage factor $\rho$ in the simulation and the shrinkage factor $\rho$ in theoretic. It indicates that the shrinkage factor is predictable in the external attacker's view.

\begin{table}[htbp]

\begin{center}

\caption{Comparison between $\rho$ in the simulation and $\rho$ in theoretic}
\label{tab1}
\setlength{\tabcolsep}{2.5mm}{
\begin{tabular}{c|ccc} \toprule
$\beta_1'$  &  $\rho$ in the simulation  &  $\rho$ in theoretic  &  Error  \\ \hline
0.25  & 0.81995  & 0.82000  & 0.00597\% \\
0.29  & 0.79482  & 0.79479  & -0.00461\%  \\
0.33  & 0.76714  & 0.76718  & 0.00528\%  \\
0.37  & 0.73471  & 0.73478  & 0.00975\% \\
0.41  & 0.69363  & 0.69336  & -0.03813\%\\
0.45  & 0.63485  & 0.63431  & -0.08428\% \\
\bottomrule

\end{tabular}}

\end{center}

\end{table}

\section{Existence of the external attacker}

In this section, according to the results in former section, we will explain why the external attacker chooses to join this system after the internal attacker's attack. In this section, the internal attacker will not change his strategy. And the external attacker launches a selfish mining attack.

\subsection{Theoretical Analysis}
Suppose that the external attacker is looking for a target cryptocurrency to attack. The external attacker is also tending to launch a selfish mining attack to the target cryptocurrency is he finds one. The external attacker infers the target system's computational power $\alpha_{target}$ through the increasing speed of the system's main chain and its mining difficulty. Meanwhile, we make the assumption that the external attacker considers all the miners in her target system honest.

The external attacker, with the computational power $\beta_2$ and the target system with inferred computational power $\alpha_{target}$ construct the multi-attacker system with the total computational power $\beta_2+\alpha_{target}$. Similarly. we normalize the computational power of this multi-attacker system. The fraction of computational power of the external attacker is: $\beta'' = \frac{\beta_2}{\beta_2+\alpha_{target}}$, and the fraction of computational power of the honest miner in the target system is: $\alpha'' = \frac{\alpha_{target}}{\beta_2+\alpha_{target}}$. Similar to the former section, we can calculate the probability that the external attacker is at a certain state:
\begin{equation}
\left\{
\begin{array}{lr}
Pr(S_{\beta_2} = 0) = \frac{\beta''-2\beta''^2}{\beta''(2\beta''^3 - 4\beta''^2 + 1)},&\\
Pr(S_{\beta_2} = 0') = \frac{\alpha''(\beta''-2\beta''^2)}{2\beta''^3 - 4\beta''^2 + 1},&\\
Pr(S_{\beta_2} = 1) = \frac{\beta''-2\beta''^2}{2\beta''^3 - 4\beta''^2 + 1},&\\
Pr(S_{\beta_2} = 2) = \frac{\beta''(\beta''-2\beta''^2)}{\alpha''(2\beta''^3 - 4\beta''^2 + 1)},&\\
Pr(S_{\beta_2} \ge 3) = \frac{\beta''^2}{\alpha''(\alpha''-\beta'')}\frac{\beta''-2\beta''^2}{2\beta''^3 - 4\beta''^2 + 1},&\\
\end{array}
\right.
\end{equation}

The expected revenue of the external attacker is: 
\begin{equation}
\begin {split}
&R_{ex}=\frac{2\beta''\alpha''(\beta''-2\beta''^2)}{2\beta''^3 - 4\beta''^2 + 1} +  \frac{(\tau_2-\beta'')\alpha''(\beta''-2\beta''^2)}{2\beta''^3 - 4\beta''^2 + 1} +\\ &\frac{2\beta''(\beta''-2\beta''^2)}{(2\beta''^3 - 4\beta''^2 + 1)}+\frac{\beta''^2}{(\alpha''-\beta'')}\frac{\beta''-2\beta''^2}{2\beta''^3 - 4\beta''^2 + 1}
\end{split}
\end{equation}

The expected revenue of honest miner in the multi-attacker system is:
\begin{equation}
\begin {split}
R_{others}= &\frac{(\tau_2-\beta'')\alpha''(\beta''-2\beta''^2)}{2\beta''^3 - 4\beta''^2 + 1}+  \frac{2(1-\tau_2)\alpha''(\beta''-2\beta''^2)}{2\beta''^3 - 4\beta''^2 + 1}+ \\
&\frac{\alpha''(\beta''-2\beta''^2)}{\beta''(2\beta''^3 - 4\beta''^2 + 1)}
\end{split}
\end{equation}
\begin{lemma}\label{lm4}
The external attacker will launch selfish mining attack to the target system if $\alpha_{target} <\frac{\beta_2(2-\beta_2-\tau_2)}{1-\tau_2}$. And in terms of parameter $\gamma_2$, the computational power of the external attacker should satisfy:
$\beta_2 > \frac{\alpha_{target}(1-\gamma_2)}{2-\gamma_2}$
\end{lemma}

\begin{IEEEproof}
As is indicated in many works, the aim of a selfish miner is to increase his or her relative revenue. In the multi-attacker system with the external attacker, his aim is $\frac{R_{ex}}{R_{ex}+R_{others}} > \beta''$. With $\alpha'' = \frac{\alpha_{target}}{\beta_2+\alpha_{target}}$ and $\beta'' = \frac{\beta_2}{\beta_2+\alpha_{target}}$, we can derive the inequality $\alpha_{target} <\frac{\beta_2(2-\beta_2-\tau_2)}{1-\tau_2}$. With the relationship: $\tau_2 = \beta'' + \gamma_2 \alpha''$ and the fact that $\gamma_2$ can be considered as a constant in a specific cryptocurrency system for the external attacker, we derive $\beta_2 > \frac{\alpha_{target}(1-\gamma_2)}{2-\gamma_2}$.
\end{IEEEproof}
\begin{lemma}\label{lm5}
The external attacker will launch the selfish mining attack to the original system which consists of internal attacker with computational power $\beta_1$ and the honest miner with computational power $\alpha$ after the internal attacker's attack if the external attacker's computational power satisfies:
\begin{equation}
\begin{split}
    &(\alpha + \beta_1)(\alpha'+ \frac{\beta'(\beta'-2\beta'^2)}{2\beta'^3 - 4\beta'^2 + 1} + \frac{\beta'\alpha'(\beta'-2\beta'^2)}{2\beta'^3 - 4\beta'^2 + 1})\\
    &< \frac{\beta_2(2-\beta_2-\tau_2)}{1-\tau_2} \le (\beta_1 + \alpha)
    \end{split}
\end{equation}

\end{lemma}
\begin{IEEEproof}
With Lemma \ref{lm4}, we know that if $ \frac{\beta_2(2-\beta_2-\tau_2)}{1-\tau_2} > (\beta_1 + \alpha)$, the external attacker would start the selfish mining attack regardless whether the internal attacker has launched an attack or not. The external attacker has already set the original system as the target before the internal attacker launches attack. Similarly, if  $ (\alpha + \beta_1)(\alpha'+ \frac{\beta'(\beta'-2\beta'^2)}{2\beta'^3 - 4\beta'^2 + 1} + \frac{\beta'\alpha'(\beta'-2\beta'^2)}{2\beta'^3 - 4\beta'^2 + 1}) > \frac{\beta_2(2-\beta_2-\tau_2)}{1-\tau_2}$, from the external attacker's perspective, even if the internal attacker launches selfish mining attack and weakens the computational power of the original system, the external attacker's computation power is still not large enough. One special case is that $ (\alpha + \beta_1)(\alpha'+ \frac{\beta'(\beta'-2\beta'^2)}{2\beta'^3 - 4\beta'^2 + 1} + \frac{\beta'\alpha'(\beta'-2\beta'^2)}{2\beta'^3 - 4\beta'^2 + 1}) = \frac{\beta_2(2-\beta_2-\tau_2)}{1-\tau_2}$. In this case, the external attacker's expected revenue in the multi-attacker system is equal to honest mining. This revenue is not large enough to motivate the external attacker to join the cryptocurrency system. 
\end{IEEEproof}
If we rewrite the inequality in terms of $\gamma_2$:
\begin{equation}
\begin{split}
   &\frac{1-\gamma_2}{2-\gamma_2}(\alpha + \beta_1)(\alpha'+ \frac{\beta'(\beta'-2\beta'^2)}{2\beta'^3 - 4\beta'^2 + 1} + \frac{\beta'\alpha'(\beta'-2\beta'^2)}{2\beta'^3 - 4\beta'^2 + 1}) \\
   &< \beta_2 \le \frac{1-\gamma_2}{2-\gamma_2}(\beta_1 + \alpha)
\end{split}
\end{equation}

We can derive the upper bound and the lower bound of the external attacker's computational power.

\subsection{Quantitative Analysis and Simulation}
\begin{figure}[htbp]

		\begin{subfigure}
			\centering
			\includegraphics[width=3in]{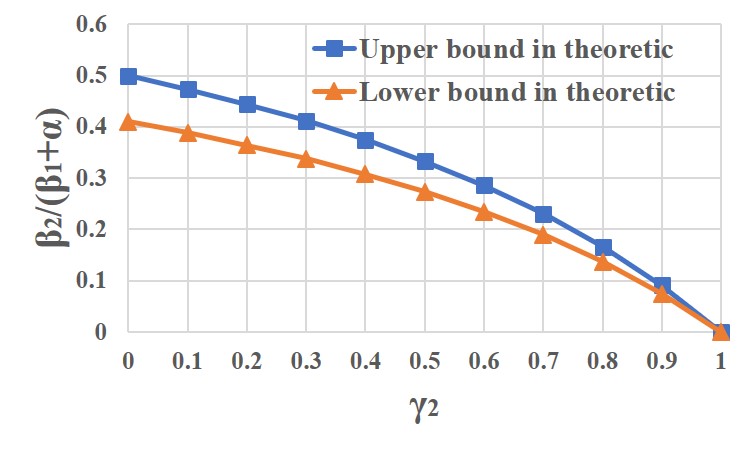}
			\caption*{(a) $\beta' = 0.25$}
		\end{subfigure}
		\begin{subfigure}
			\centering
			\includegraphics[width=3in]{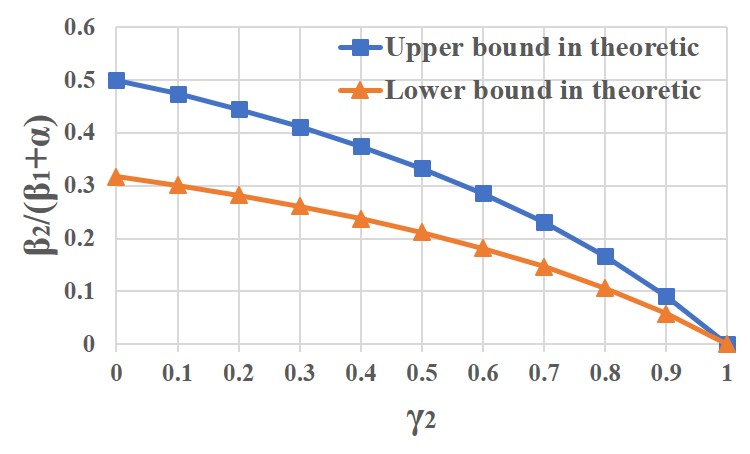}
			\caption*{(b)$\beta' = 0.45$}
		\end{subfigure}
		
		\caption{The upper bound and lower bound of the external attacker's computational power.}
		\label{bound}
\end{figure}

In this section, we first consider two cases: In the original system, the parameter $\beta'$ equals to 0.25 and 0.45. According to Table \ref{tab1}, the theoretical shrinkage factor in these two cases is 0.82 and 0.63431 respectively. In Fig. \ref{bound}(a) $\beta'$ equals 0.25, and in Fig. \ref{bound}(b) $\beta'$ equals 0.45. In these two figures, Y-axis represents ratio of the external attacker's computational power to the original system's computational power. Compared with Fig. \ref{bound}(b), in Fig. \ref{bound}(a) the gap between the upper bound and the lower bound is narrower.

For each $\beta'$, there exists a specific lower bound for the external attacker. In the external attacker's perspective, as long as the ratio of his computational power to the original system's computational power is above the curve of lower bound, he can always gain extra revenue. This is a reasonable but unrealistic expectation since the external consider the original system as an honest entity.

To explain why the external attacker's expectation is an unrealistic one, we simulate two specific cases. In the first case, the parameter $\beta'$ equals to 0.25 which results in $\rho = 0.82$. We set the computational power of the external attacker as $\frac{1}{2}\times 0.82 \times \alpha_{target}$ so that it will always satisfy the inequality $\frac{1-\gamma_2}{2-\gamma_2}(\alpha + \beta_1)(\alpha'+ \frac{\beta'(\beta'-2\beta'^2)}{2\beta'^3 - 4\beta'^2 + 1} + \frac{\beta'\alpha'(\beta'-2\beta'^2)}{2\beta'^3 - 4\beta'^2 + 1}) < \beta_2$ when $\beta' = 0.25$. After normalization, we get the computational power of the honest miner, the internal attacker and the external attacker in the first case which is $\alpha = \frac{3}{5.64}$, $\beta_1 = \frac{1}{5.64}$ and $\beta_2 = \frac{1.64}{5.64}$ respectively. In the second case, the parameter $\beta'$ equals to 0.45 which results in $\rho \approx 0.63431$. The external attacker's computational power is set as $\frac{1}{2}\times 0.63431 \times \alpha_{target}$. After normalization, we get the computational power of the honest miner, the internal attacker and the external attacker in the second case which is $\alpha \approx \frac{11}{26.34}$, $\beta_1 \approx \frac{9}{26.34}$ and $\beta_2 \approx \frac{6.34}{26.34}$ respectively.
We use a Monte Carlo method to generate a blockchain with the height $10^6$ blocks and iterate for 100 times.

\begin{figure}[htbp]

		\begin{subfigure}
			\centering
			\includegraphics[width=3in]{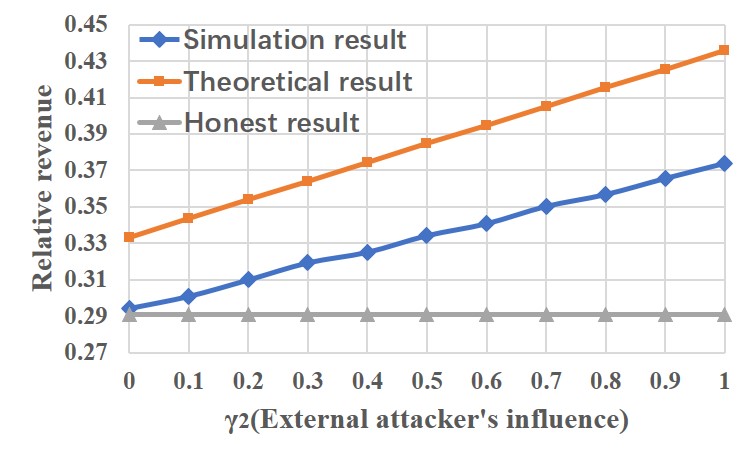}
			\caption*{(a)$\beta' = 0.25$}
		\end{subfigure}
		\begin{subfigure}
			\centering
			\includegraphics[width=3in]{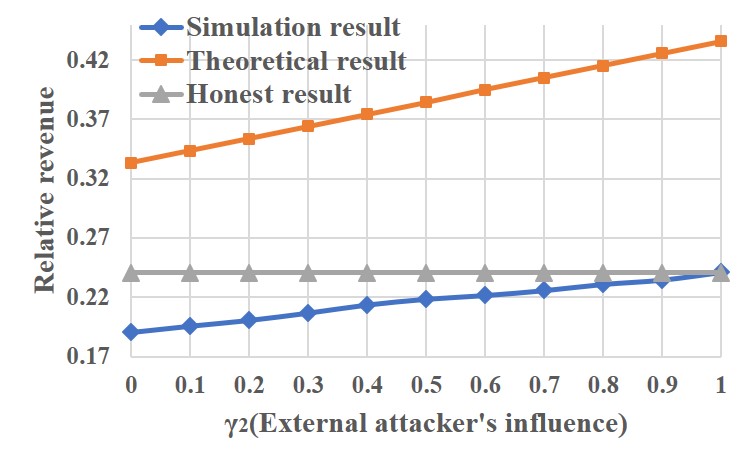}
			\caption*{(b)$\beta' = 0.45$}
		\end{subfigure}
		
		\caption{Comparison between external's relative revenue in theoretic and in simulation}
		\label{comp}
\end{figure}

Fig. \ref{comp}(a) shows the simulation result of the first case and demonstrates that the external attacker will not earn as much revenue as he expected. Here, we define relative revenue as $\frac{R_{ex}}{R_{ex} + R_{others}}$ where $R_{ex}$ is the block reward of the external attacker, and $R_{others}$ represents the block reward earned by the internal attacker and the honest miner. The horizontal curve is the relative revenue that the external attacker will receive if he mines honestly. In the simulation with $\beta' = 0.25$, the relative revenue of the external attacker is still higher than the relative revenue of the external attacker if he mines honestly. 

Fig. \ref{comp}(b) demonstrates that when $\beta' = 0.45$, the external attacker's relative revenue is even less than that of honest mining. Note that, the theoretical results predicted by the external attacker in the two cases are the same. From the external attacker's perspective, the ratio of his computational power to the original system's computational power is $\frac{1}{2}$ in both 2 cases.

Table \ref{tab2} shows the difference between the relative revenue in external's expectation and that in the simulation when $\beta' = 0.25$. The result suggests that the external attacker cannot predict his block reward precisely before he launches the attack. The higher influence factor $\gamma_2$ is in the external attacker's prediction, the larger the gap is between the relative revenue in external's expectation and that in the simulation. In this case, an error which is less than $15\%$ is acceptable for the external attacker since he can still earn some extra revenue by launching an attack.

\begin{table}[htbp]

\begin{center}
\caption{Comparison between external attacker's relative revenue in the simulation and that in theoretic when $\beta' = 0.25$.}
\label{tab2}
\setlength{\tabcolsep}{2.5mm}{
\begin{tabular}{c|ccc} \toprule
$\gamma_2$  &  Theoretical result  & Simulation result  &  Error  \\ \hline
0  & 0.33333  & 0.29441  & -11.67734\% \\
0.2  & 0.35384  & 0.30996  & -12.40035\%  \\
0.4  & 0.37435  & 0.32504  & -13.17319\%  \\
0.6  & 0.39487  & 0.34063  & -13.73518\% \\
0.8  & 0.41538  & 0.35655  & -14.16162\%\\
1.0  & 0.43589  & 0.37355  & -14.30192\% \\
\bottomrule

\end{tabular}}

\end{center}

\end{table}

Table \ref{tab3} shows the difference between the relative revenue in external's expectation and that in the simulation when $\beta' = 0.45$. The error becomes unacceptable since it is lar/ower of the internal attacker can seriously affect the decision of the external attacker. When the internal attacker has a high computational power which results in a larger $\beta'$ and a smaller shrinkage factor $\rho$, the external attacker may launch an attack against the original system with the computational power which is far from enough.

\begin{table}[htbp]

\begin{center}
\caption{Comparison between external attacker's relative revenue in the simulation and that in theoretic when $\beta' = 0.45$.}
\label{tab3}
\setlength{\tabcolsep}{2.5mm}{
\begin{tabular}{c|ccc} \toprule
$\gamma_2$  &  Theoretical result  & Simulation result  &  Error  \\ \hline
0  & 0.33333  & 0.19036  & -42.89060\% \\
0.2  & 0.35384  & 0.20056  & -43.31764\%  \\
0.4  & 0.37435  & 0.21361  & -42.93750\%  \\
0.6  & 0.39487  & 0.22163  & -43.871046\% \\
0.8  & 0.41538  & 0.23104  & -44.376730\%\\
1.0  & 0.43589  & 0.24120  & -44.664646\% \\
\bottomrule

\end{tabular}}

\end{center}
\end{table}

\section {Multi-Attacker System}

The former section has indicated that the external attacker always overestimates his relative revenue before he launches an attack against the original system. The gap between the estimation of the relative revenue and the relative revenue in real case becomes more and more unacceptable for the external attacker when the internal attacker's computational power becomes larger.

Similar to the external attacker, after the external attacker's attack, the internal attacker's relative revenue cannot meet his expectation either. In this section, we will demonstrate how the internal attacker notices the existence of the external attacker. We also infers the reasons that both external attacker and internal attack's relative revenue do not meet their expectation. Further more, we presents some countermeasures for the internal attacker after he notices the external attacker.

\subsection{The External Attacker's Influence}
We define three stage so as to demonstrate how much the internal attacker lose and how the internal attacker notices the existence of the external attacker.
\begin{itemize}
    \item[1)] Stage one: In stage one, the external attacker has not launched an selfish mining attack against the original system. The relative revenue of the internal attacker corresponds with the internal attacker's expectation. 
    \item[2)] Stage two: In stage two, the external attacker launches the attack. The internal attacker notices the existence of external computational power, but he has not been aware that the external computational power is an attacker. According to the probability that the external computational power finds a block, the internal attacker can easily estimate the entity's computational power. After normalization, the internal attacker's computational power is $\beta_1$ in multi-attacker system. The internal attacker makes a new expectation of his relative revenue according to the value of $\beta_1$.
    \item[3)] Stage three: In stage three, the internal attacker notices that his relative revenue is less than his expectation. This fact indicates that the external computational power is an attacker as well.
\end{itemize}

We conduct two specific cases to better illustrate the process for the internal attacker to notice the existence of the external attacker. To accord with the simulations in former sections, in the first case, parameter $\beta' = 0.25$ which results in $\rho = 0.82$. The computational power of the external attacker is $\frac{1}{2}\times \rho\times \alpha_{target}$ so that after normalization, $\alpha = \frac{3}{5.64}$, $\beta_1 = \frac{1}{5.64}$ and $\beta_2 = \frac{1.64}{5.64}$. In the second case, $\beta' = 0.45$ and $\rho \approx 0.63431$. The external attacker's computational power is $\frac{1}{2}\times \rho \times \alpha_{target}$ so that $\alpha \approx \frac{11}{26.34}$, $\beta_1 \approx \frac{9}{26.34}$ and $\beta_2 \approx \frac{6.34}{26.34}$. In the simulation of both cases, we use Monte Carlo method to generate a blockchain with $10^6$ blocks for 100 times.

\begin{figure}[htbp]

		\begin{subfigure}
			\centering
			\includegraphics[width=3in]{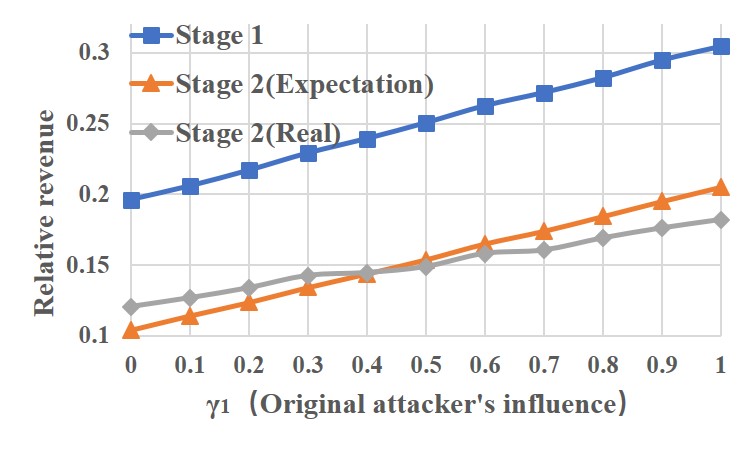}
			\caption*{(a)$\beta' = 0.25$ }
		\end{subfigure}
		\begin{subfigure}
			\centering
			\includegraphics[width=3in]{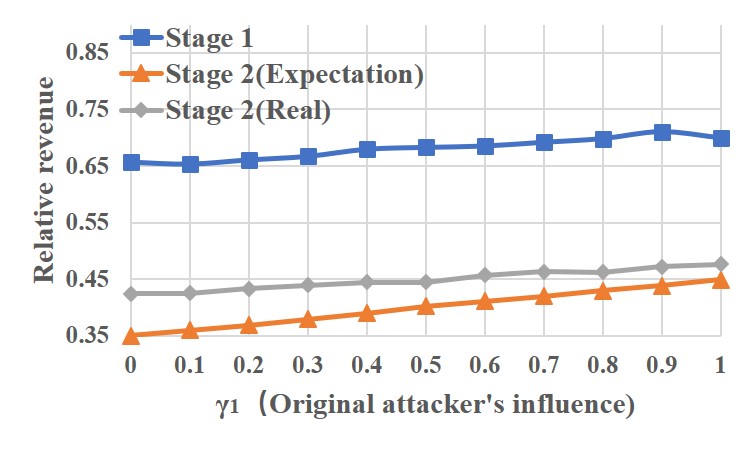}
			\caption*{(b)$\beta' = 0.45$}
		\end{subfigure}
		
		\caption{Relative revenue of the internal attacker in simulation and in his expectation }
		\label{loss}
\end{figure}

Fig. \ref{loss}(a) demonstrates the relative revenue of the internal attacker when $\beta' = 0.25$ in state one and stage two. In stage one, the theoretical result is approximately equivalent to the simulation result so that the internal attacker can precisely calculate his relative revenue in stage one. In stage two, the internal attacker has not noticed that the external computational power is also an attacker. He calculates a new expectation based on his new computational power $\beta_2$. But at stage two, his expected relative revenue does not hold the relative revenue in the real case. Especially when $\gamma_1 > 0.5$, the expected relative revenue exceeds his relative revenue in the real case.

Fig. \ref{loss}(b) demonstrates the relative revenue of the internal attacker when $\beta' = 0.45$ in state one and stage two. In this case, relative revenue of the internal attacker exceeds internal attacker's expectation.

Table \ref{tab4} and Table \ref{tab5} show that, in both two cases, there is a significant reduction of the original's relative revenue from stage one to stage two. The reduction of relative revenue let the internal attacker be aware of the existence of the external computational power. In stage two, the difference between internal attacker's relative revenue in real case and that in expectation let the internal attacker be aware that the external computational power is an attacker.

\begin{table}[htbp]
\begin{center}

\caption{Comparison of the internal attacker's relative revenues when $\beta' = 0.25$.}
\label{tab4}
\setlength{\tabcolsep}{1.75mm}{
\begin{tabular}{c|ccc} \toprule
$\gamma_1$  &  Stage two(real)  &  Stage two(Expectation)  &  Stage one  \\ \hline
0  & 0.120  &0.104(-13.672\%)  & 0.196(62.65\%) \\
0.2  & 0.133  & 0.123(-7.632\%)  & 0.216(62.051\%)  \\
0.4  & 0.144  & 0.143(-0.459\%)  & 0.239(65.584\%)  \\
0.6  & 0.158  & 0.165(4.412\%)  & 0.262(66.136\%) \\
0.8  & 0.168  & 0.184(9.121\%)  & 0.282(67.163\%)\\
1.0  & 0.181  & 0.205(12.751\%)  & 0.304(67.461\%) \\
\bottomrule

\end{tabular}}

\end{center}
\end{table}

\begin{table}[htbp]
\begin{center}
\caption{Comparison of the internal attacker's relative revenues when $\beta' = 0.45$.}
\label{tab5}
\setlength{\tabcolsep}{1.75mm}{
\begin{tabular}{c|ccc} \toprule
$\gamma_1$  &  Stage two(real)  &  Stage two(Expectation)  &  Stage one \\ \hline
0  & 0.424  &0.350(-17.413\%)  & 0.657(54.848\%) \\
0.2  & 0.433  & 0.368(-14.880\%)  & 0.653(52.685\%)  \\
0.4  & 0.444  & 0.390(-12.188\%)  & 0.679(52.908\%)\\
0.6  & 0.457  & 0.411(-10.140\%)  & 0.685(49.846\%) \\
0.8  & 0.462  & 0.430(-7.009\%)  & 0.698(50.935\%)\\
1.0  & 0.476  & 0.449(-5.663\%)  & 0.699(46.869\%) \\
\bottomrule

\end{tabular}}

\end{center}

\end{table}

\subsection{Reasons for Both Attackers' Loss}
Both the external attacker and the internal attacker's expectations of their RR differs from their RR in the real case. Even internal attacker's expected RR in stage two is sometimes higher than his real RR in state two, there is a significant reduction of his RR from state one to stage two.

The external attacker and the internal attacker's wrong expectations of their RR result from the fact that they use the basic attacking model with a single attacker to predict their RR. Some cases in the multi-attacker system are not considered in attacking models with a single attacker.

\subsubsection{Competition between attackers}
Competitions, or forks in the multi-attacker system not only exist between one attacker(or both two attackers) and the honest miner, but also exist between the external attacker and the internal attacker. We define the competitions which include the honest miner as Type one and the competitions between internal attacker and external attackers as Type two.

Two type of competitions differ in how they are generated. Type one competition results from the action Match of the attackers and Type two competition results from action Override of two attackers. We present a simple case study to make it clear.

\begin{figure}[htbp]
    \centering
    \includegraphics[width = 3in]{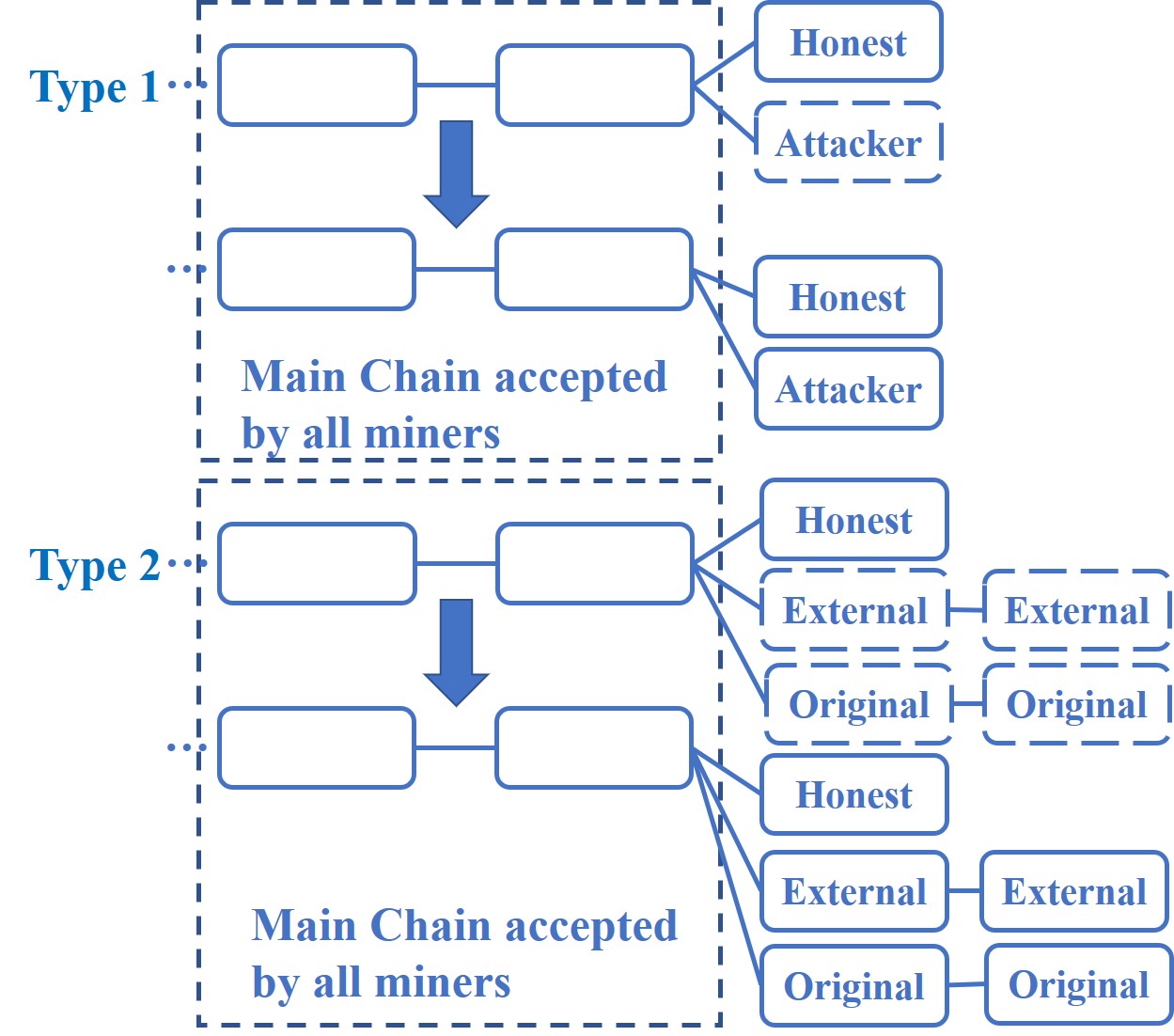}
    \caption{Two type of competitions}
    \label{2com}
\end{figure}

Fig. \ref{2com} demonstrates two specific cases of type one competition and type two competition respectively. The dash line in the figure means that the block is unreleased. In type one competition, the honest miner release a newly found block while one attacker(Either the internal attacker or external attacker) is holding a unreleased block. After receiving the block found by honest miner, The attacker takes action Match to generate a fork in the blockchain. In type two competition, the honest miner release a block while both two attackers have two unreleased blocks. Both attackers take action Override. Neither of them intends to form a competition in the blockchain, but a unexpected competition is generated.

Type two competitions are unexpected competitions. These competitions waste the attackers' computational power since only one of the attackers' released chain can be eventually accepted as the main chain by all miners. The blocks on another unaccepted chain are staled.

\subsubsection{Action Override made by another attacker}
Unexpected competitions waste attacker's computational power, but action Override made by another attacker harms the attacker more. In mining attacker models with a single attacker, it is unnecessary for the attacker to consider the risk that other miner will override his released chain. But in our model, both external and internal attacker have to worry about this risk.

\begin{figure}[htbp]
    \centering
    \includegraphics[width = 3in]{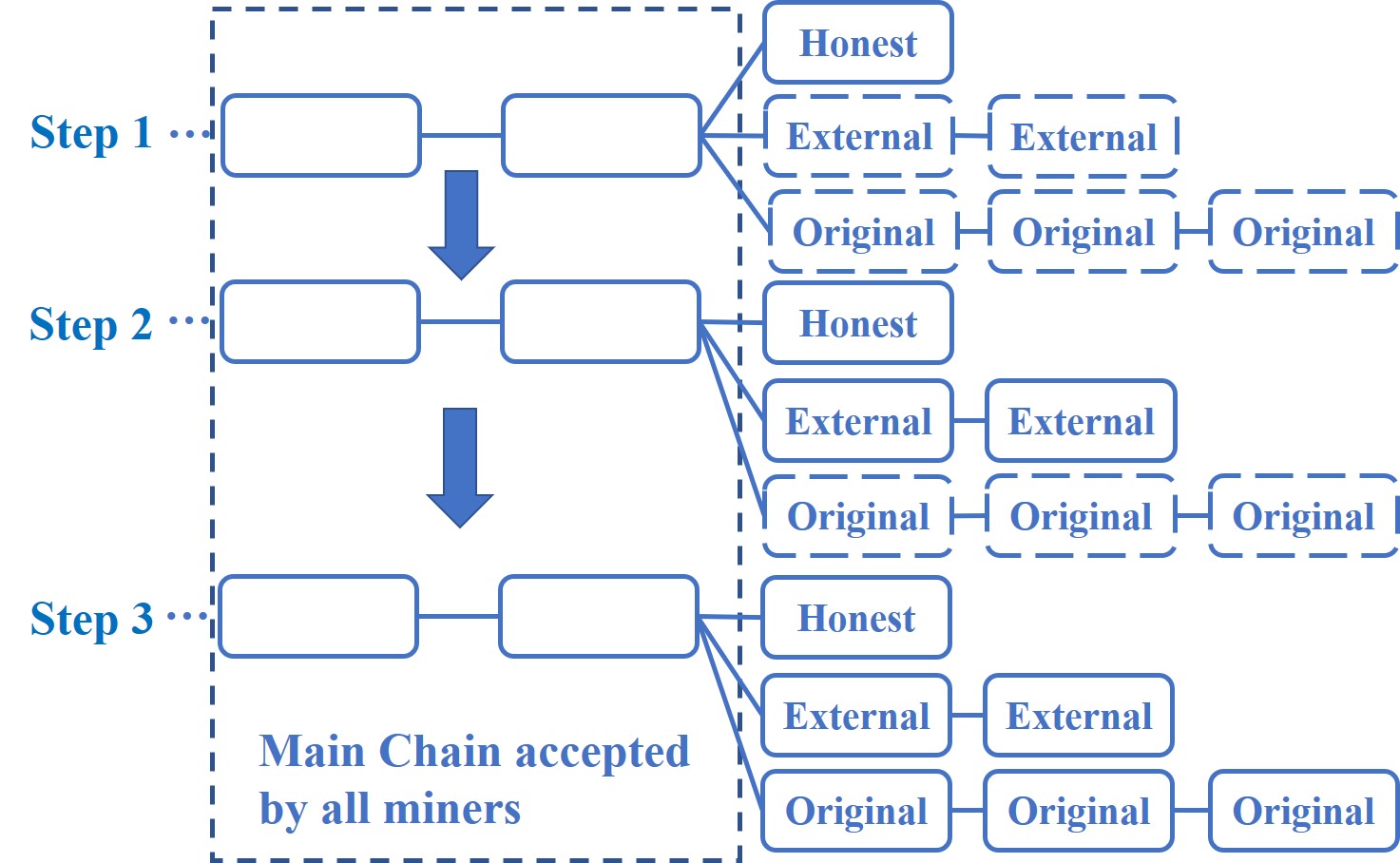}
    \caption{The auction between the attackers.}
    \label{auction}
\end{figure}

Action Override of another attacker can be described as the auction between external attacker and the internal attacker. Fig. \ref{auction} is the simplest case in which the auction between the attackers occurs. The dash line in the figure means that the block is unreleased. Suppose in step one, the external attacker holds two unreleased blocks and the internal attacker holds three unreleased blocks. The honest miner releases a newly found block. Suppose both attackers' mining strategies are selfish mining. Then, in step two, the external attacker takes the action Override and the internal attacker takes the action hold since he has not received the blocks released by the external attacker yet. In step 3, the original received the blocks released by the external attacker. So he takes action Override as well. 

In this case study, two of the external attacker's blocks are staled and one of the honest miner's block is staled. The external attacker's loss is more than the loss of honest miner.

\subsubsection{Support from honest miner is split}
Unexpected competitions and auctions are two reasons that attackers wrongly estimate their relative revenue. Besides, attackers' influence parameters $\gamma_1$ and $\gamma_2$ are always overestimated by themselves. 

When attackers estimate of $\gamma_1$ and $\gamma_2$ in former sections, they have not realized the existence of another attacker yet. Suppose the fraction of computational power of the honest miner affected by the attackers is $\gamma_h$. internal attacker and external attacker will estimate his influence rate as $\gamma_1 = \gamma_h$ and $\gamma_2 = \gamma_h$. 

Apparently, attacker's overestimate of their influence rate is another factor that causes attackers' wrongly estimation of their RR. 

In our model, we calculate attacker's real influence rate $\gamma_1'$ and $\gamma_2'$ based on the following three steps:
\begin{itemize}
    \item Denote the fraction of computational power of the honest miner affected by the attackers as $\gamma_h$. If there is a type two competition in the blockchain, $\gamma_h = 1$.
    \item If the competition is a type two competition, $\gamma_1' = \frac{\beta_1}{\beta_2 + \beta_1}$, and $\gamma_2' = \frac{\beta_2}{\beta_2 + \beta_1}$
    \item If the competition is a type one competition, denote the computational power of all attackers involved in the competition as $\beta$. If internal attacker is in the competition, $\gamma_1' = \gamma_h \times \frac{\beta_1}{\beta}$ and $\gamma_2' = \gamma_h - \gamma_1'$. Otherwise, $\gamma_1' = 0$ and $\gamma_2' = \gamma_h$.
\end{itemize}

\subsection{Catfish Effect in Multi-attacker System}
Due to the existence of the external attacker, the internal attacker need to seek a better attacking strategy.

In section \uppercase\expandafter{\romannumeral6}. B, we propose three reasons for the wrong expectation of the internal attacker and the external attacker. They are unexpected competitions, auctions between attackers and overestimation of influence factor.

For the internal attacker, auctions between attackers are inevitable because the external attacker's state is unavailable to the internal attacker. 

But the influence factor of the internal attacker can be increased. internal attacker's overestimate of his influence factor $\gamma_1$ only exists in the cases that both attackers are involve the competition. Note that a great part of these cases are unexpected competitions.

The countermeasures of the internal attacker can be considered as a mining strategy which reduces unexpected competitions and wastes other miner's computational power.

\subsubsection{Mining Honestly}
An interesting fact is that, mining honestly can be considered as an effective counter method. As mentioned in Section \uppercase\expandafter{\romannumeral5}. A, the upper bound of the computational power of the external attacker is $\frac{1-\gamma_2}{2-\gamma_2}(\beta_1 + \alpha)$. This upper bound ensures that even in the worst case in which the external's computational power reaches its upper bound, the internal attacker can earn as much revenue as he deserves.

The internal attacker can avoid unexpected competitions by mining honestly. But he fails to waste the other miner's computational power.

\subsubsection{Partial initiative release}

We propose a new mining attack strategy: partial initiative release(PIR). PIR is designed for the mining attacker model with two or more attackers. Similar to selfish mining, PIR's state transition is based on a state machine.

\begin{figure}[htbp]
    \centering
    \includegraphics[width = 3in]{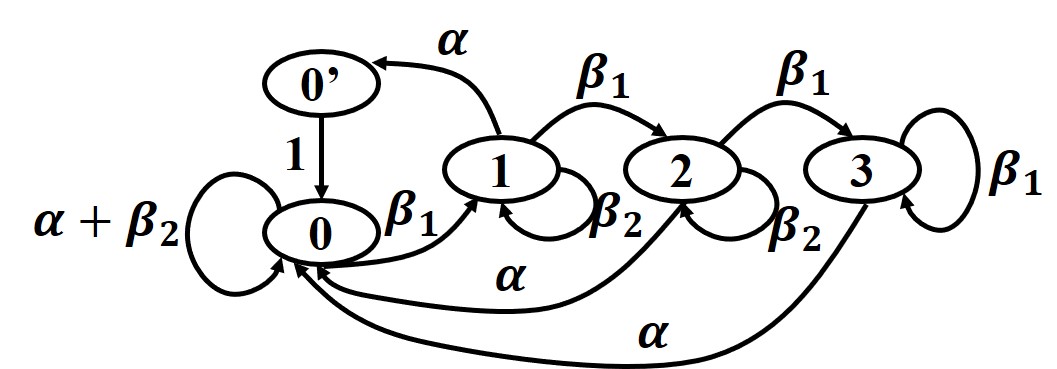}
    \caption{The state machine of $PIR_3$.}
    \label{pir3}
\end{figure}

PIR is a strategy set which consists of \{$PIR_1,PIR_2,\cdots,PIR_n,\cdots$\}. Fig. \ref{pir3} is the state machine of $PIR_3$. By demonstrating how $PIR_3$ works, we explain why PIR is suitable for multi-attacker system.

When the internal attacker is at state $S=i$, a block released by honest miner results in the internal attacker's state transition from $S=i$ to $S = i-1$. A block found by internal attacker himself results in the internal attacker's state transition from $S=i$ to $S = i+1$. In one round of multi-attacker system, the probability that an honest miner finds and releases a block is $\alpha$, and the probability that the original miner finds a block is $\beta_1$. When the external attacker finds the block in one round with probability $\beta_2$, the internal attacker's state does not change. 

Different from state machine of selfish mining, the number of states is finite. In $PIR_3$, the max state is three. When the internal attacker's state is $S=3$ and he finds the next block, he will initiative release a block to ensure that his state does not exceed three. When the honest miner releases a block, the internal attacker will release all his three unreleased blocks. Since the internal attacker will take the action Release in some specific situation, we named this strategy as Partial initiative release.

$PIR_3$ cannot completely avoid unexpected competitions. For example, when the internal attacker takes action Override when $S = 2$ and the external attacker happens to take action Override, an unexpected competition shows up.

But when $S = 3$ and the honest miner releases a block, initiative releasing all three blocks can lower the probability of unexpected competitions. We explain how initiative releasing all blocks lowers the probability of unexpected competitions by three cases:
\begin{itemize}
    \item[1)] External attacker's state is lower than three. The external attacker will take action Adopt which ensures that the released blocks of internal attacker can be accepted by all miners.
    \item[2)] External attacker's state is equal to three. the external attacker will take action Match. Consider the case that internal attacker takes action Hold instead of releasing all three blocks. An unexpected competition will occur if the honest miner finds and releases another block. In unexpected competition, the two attackers take action Override at roughly the same time so that the honest miner's support is split. But when internal attacker initiative releases all three blocks and external attacker takes action Match, the blocks of internal attacker are released before external attacker's. So most of the honest miners receive and accept internal attacker's blocks. In this case, internal attacker gains more support from the honest miner.
    \item[3)] External attacker's state is greater than three. As is indicated in the former section, in this case, action Override made by external attacker is inevitable. But the external attacker's Override prevents the internal attacker from wasting more computational power.
\end{itemize}

\subsection{Quantitative Analysis and Simulation}
In this section, we analyze internal attacker's countermeasures through simulations. We demonstrate when the internal attacker should take countermeasures and which countermeasure should be taken.



In the simulations of this section, we set $\beta'$ as the variable. Meanwhile $\gamma_h$ is chosen from the set $ \{0,0.25,0.5,0.75 \}$. The shrinkage factor $\rho = (\alpha'+ \frac{\beta'(\beta'-2\beta'^2)}{2\beta'^3 - 4\beta'^2 + 1} + \frac{\beta'\alpha'(\beta'-2\beta'^2)}{2\beta'^3 - 4\beta'^2 + 1})$. For a more comprehensive explanation of the problem, we simulate the cases with $\beta_2 = \frac{1-\gamma_h}{2-\gamma_h}\rho(\alpha +\beta_1)$ and $\beta_2 = \frac{1-\gamma_h}{2-\gamma_h}(\alpha +\beta_1)$. The two values of $\beta_2$ is the upper bound and lower bound respectively in Fig. \ref{bound}.Meanwhile, $\beta_1 = \beta'(\beta_1 + \alpha)$ and $\alpha + \beta_1 +\beta_2 = 1$. When $\beta_2$ is the upper bound, $\alpha = \frac{(1-\beta')(2-\gamma_h)}{(1-\gamma_h)\rho + (2-\gamma_h)}$, $\beta_1 =\frac{\beta'(2-\gamma_h)}{(1-\gamma_h)\rho + (2-\gamma_h)} $ and $\beta_2 = \frac{(1-\gamma_h)\rho}{(1-\gamma_h)\rho + (2-\gamma_h)}$. When $\beta_2$ is the lower bound, $\alpha = \frac{(1-\beta')(2-\gamma_h)}{3-2\gamma_h}$, $\beta_1 = \frac{\beta'(2-\gamma_h)}{3-2\gamma_h}$ and $\beta_2 = \frac{1-\gamma_h}{3-2\gamma_h}$.

\subsubsection{$\beta_2$ is the lower bound}
 We compare RRs of the internal attacker when he takes different mining strategies including selfish mining, honest mining and $PIR_3$. We also present the curve of $\beta_1$ according to $\beta'$ as the baseline. With $RR > \beta_1$, the internal attacker can earn extra revenue.
\begin{figure} [htbp]

\centering 
\begin{minipage}[h]{0.5\linewidth}
\subfigure[$\gamma_h = 0$]{\label{fig:subfig:a}

\includegraphics[width=1.7in]{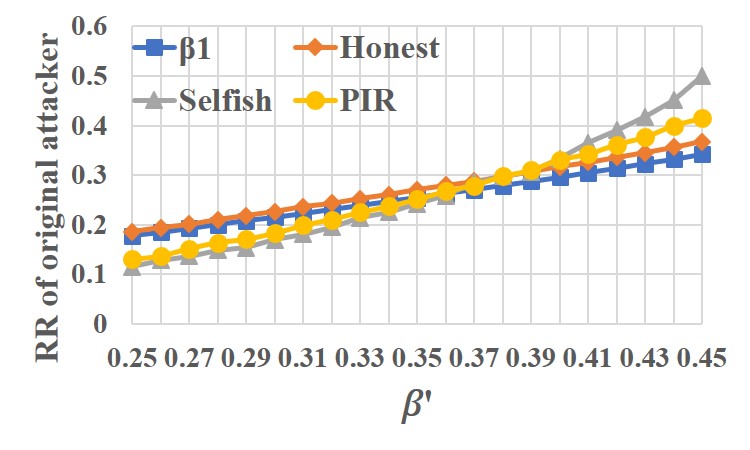}}

\subfigure[$\gamma_h = 0.25$]{\label{fig:subfig:b}

\includegraphics[width=1.7in]{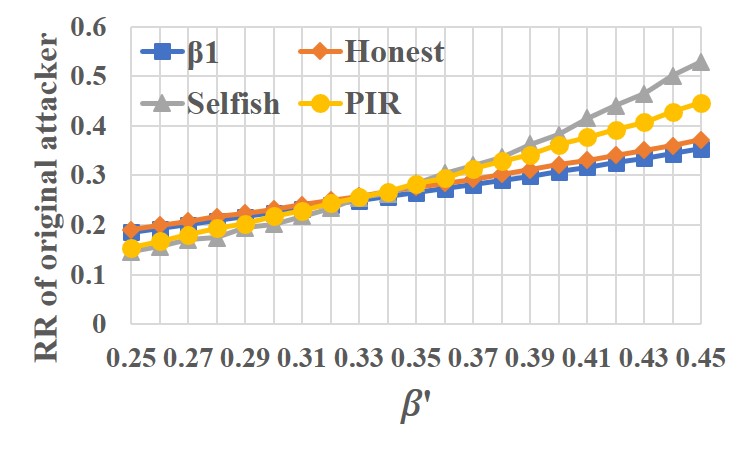}}

\end{minipage}%
\begin{minipage}[h]{0.5\linewidth}
\subfigure[$\gamma_h = 0.5$]{\label{fig:subfig:a}

\includegraphics[width=1.7in]{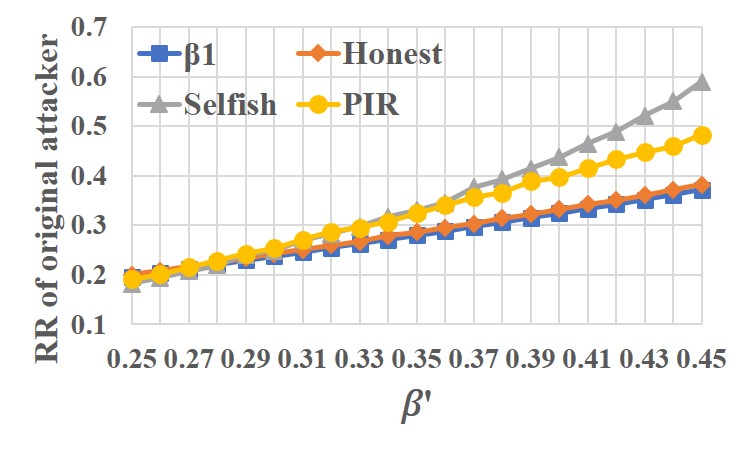}}

\subfigure[$\gamma_h = 0.75$]{\label{fig:subfig:b}

\includegraphics[width=1.7in]{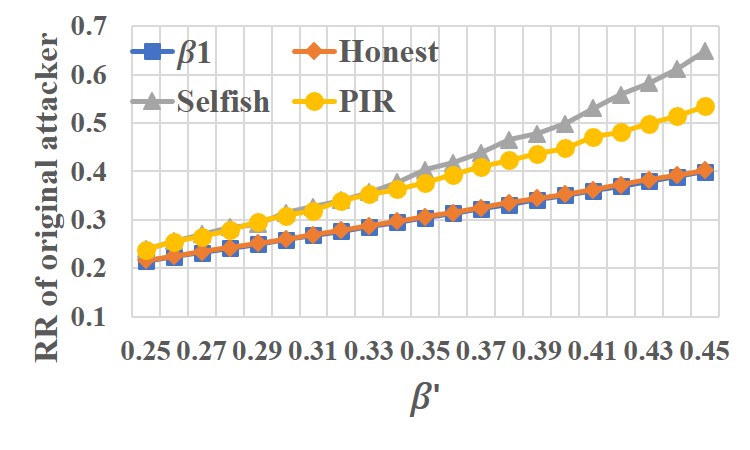}}
\end{minipage}%
\caption{Internal attacker's relative revenue when $\beta_2$ is at the lower bound.}

\label{lowb}

\end{figure}

Fig. \ref{lowb} demonstrates the internal attacker's relative revenue when he takes different mining strategies including selfish mining, honest mining and $PIR_3$. We consider $\gamma_h = 0$ and $\gamma_h = 0.25$ as low $gamma_h$ values while $\gamma_h = 0.5$ and $\gamma_h = 0.75$ as high $\gamma_h$ values. In the four cases, when $\beta'$ is high, selfish mining outperforms honest mining and $PIR_3$. This result suggests that when $\beta'$ is high, the internal attacker do not need to take any countermeasures. When $\beta'$ is low and $\gamma_h$ is low, honest mining is the best strategy and both selfish mining and $PIR_3$ is under the baseline. In the cases in which $\beta'$ is low and $\gamma_h$ is high, $PIR_3$ beats other mining strategies by a slight advantage. Table \ref{tab6} shows the best strategy among selfish mining, honest mining and $PIR_3$
\begin{table}[htbp]
\begin{center}
\caption{The best strategy of the internal attacker when $\beta_2$ is at the lower bound.}
\label{tab6}
\setlength{\tabcolsep}{2.5mm}{
\begin{tabular}{c|cc} \toprule
\diagbox{$\beta'$}{$\gamma_h$}  &  Low  &  High \\ \hline
Low  & Honest mining  & $PIR_3$  \\
High  & Selfish mining & Selfish mining   \\

\bottomrule

\end{tabular}}

\end{center}
\end{table}
\subsubsection{$\beta_2 $is the upper bound} We compare RRs of the internal attacker as well when $\beta_2 $is the upper bound.

\begin{figure} [htbp]

\centering 
\begin{minipage}[h]{0.5\linewidth}
\subfigure[$\gamma_h = 0$]{\label{fig:subfig:a}

\includegraphics[width=1.7in]{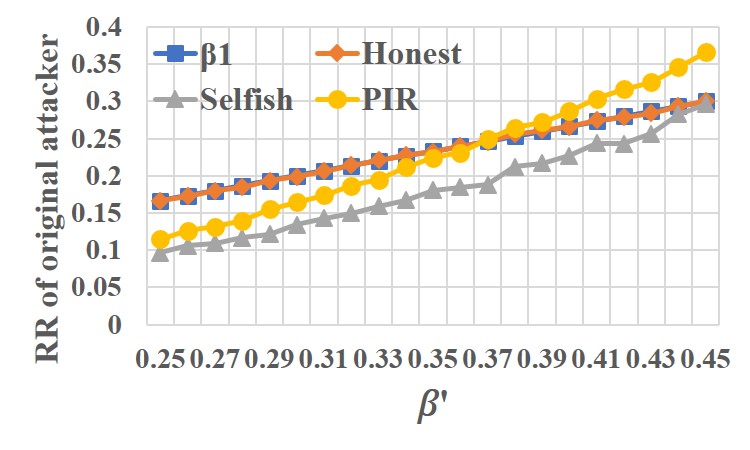}}

\subfigure[$\gamma_h = 0.25$]{\label{fig:subfig:b}

\includegraphics[width=1.7in]{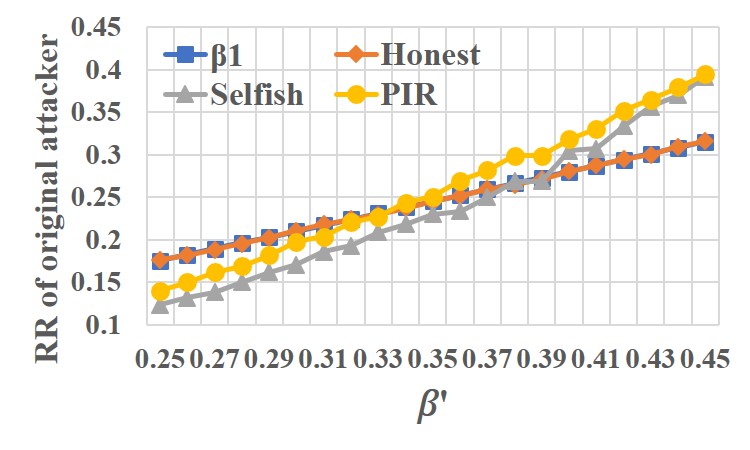}}

\end{minipage}%
\begin{minipage}[h]{0.5\linewidth}
\subfigure[$\gamma_h = 0.5$]{\label{fig:subfig:a}

\includegraphics[width=1.7in]{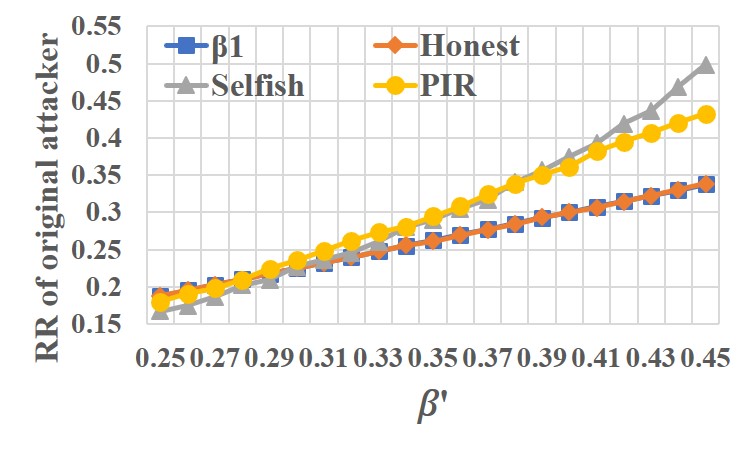}}

\subfigure[$\gamma_h = 0.75$]{\label{fig:subfig:b}

\includegraphics[width=1.7in]{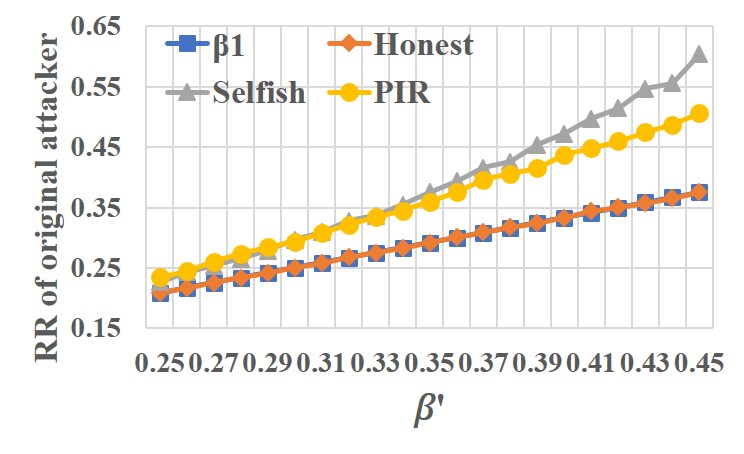}}
\end{minipage}%
\caption{Internal attacker's relative revenue when $\beta_2$ is at the upper bound}

\label{upb}

\end{figure}

Fig. \ref{upb} demonstrate the internal attacker's relative revenue when he takes different mining strategies including selfish mining, honest mining and $PIR_3$. In the four cases, note that the curve of honest mining always coincides the baseline. When $\beta_2$ is at the upper bound, internal attacker can ensure that he can earn as many block rewards as he deserves by mining honestly. Table \ref{tab7} demonstrates the best strategy among selfish mining, honest mining and $PIR_3$ in different cases when $\beta_2$ is at the upper bound. Table \ref{tab7} differs from Table \ref{tab6} when $\beta'$ is high and $\rho$ is low.
\begin{table}[htbp]
\begin{center}

\caption{The best strategy of the internal attacker when $\beta_2$ is at the upper bound.}
\label{tab7}
\setlength{\tabcolsep}{2.5mm}{
\begin{tabular}{c|cc} \toprule
\diagbox{$\beta'$}{$\gamma_h$}  &  Low  &  High \\ \hline
Low  & Honest mining  & $PIR_3$  \\
High  & $PIR_3$ & Selfish mining   \\

\bottomrule

\end{tabular}}
\end{center}
\end{table}
\section{Conclusion}




In this paper, we propose an attacking model in a proof-of-work blockchain with an internal attacker and an external attacker. Our model consists of two phase: the original system and the multi-attacker system. From our theoretic and quantitative analysis, we demonstrate the catfish effect between the internal attacker and the external attacker. The internal attacker has to improve his attacking strategy due to the threat brought by the external attacker. We propose an attacking strategy in multi-attacker system named Partial Initiative Release (PIR). An interesting fact is that, mining honestly is another countermeasure of the internal attacker. Our simulation results shows that the original can select a mining strategy among PIR, honest mining and selfish mining based on the parameter $\beta'$,$\gamma_h$ and $\beta_2$.


%




\ifCLASSOPTIONcaptionsoff
  \newpage
\fi

\end{document}